\documentclass[conference]{IEEEtran}
\IEEEoverridecommandlockouts
\usepackage{cite}
\usepackage{amsmath,amssymb,amsfonts}
\usepackage{algorithmic}
\usepackage{graphicx}
\usepackage{textcomp}
\usepackage{xcolor}
\usepackage{bm}

\usepackage[caption=false,font=footnotesize]{subfig}
\usepackage{stfloats}
\usepackage{enumerate,amsthm,}

\newtheorem{definition}{Definition}

\newtheorem{theorem}{Theorem}
\newtheorem{lemma}{Lemma}
\newtheorem{corollary}{Corollary}

\graphicspath{{figures/}}
\def\BibTeX{{\rm B\kern-.05em{\sc i\kern-.025em b}\kern-.08em
    T\kern-.1667em\lower.7ex\hbox{E}\kern-.125emX}}
\begin{document}

\title{Compressive Multidimensional Harmonic Retrieval with Prior Knowledge \\
\thanks{This work was supported by the Key Program of National Natural Science Foundation of China (No. 11833001), the National Science Fund for Distinguished Yong Scholars (No. 61625103), the Leading Talents Program of Guangdong Province (Grant 00201510), the Shenzhen Peacock program (Grant KQTD2015071715073798) and the China Scholarship Council. (Corresponding author: Zegang Ding, e-mail: z.ding@bit.edu.cn)}
}

\author{\IEEEauthorblockN{1\textsuperscript{st} Yinchuan Li}
\IEEEauthorblockA{\textit{School of Information and Electronics} \\
\textit{Beijing Institute of Technology}\\
Beijing, China \\
yinchuan.li.cn@gmail.com}
\and
\IEEEauthorblockN{2\textsuperscript{nd} Xu~Zhang}
\IEEEauthorblockA{\textit{School of Information and Electronics} \\
\textit{Beijing Institute of Technology}\\
Beijing, China \\
xu.zhang.bit@gmail.com}
\and
\IEEEauthorblockN{3\textsuperscript{rd} Zegang~Ding}
\IEEEauthorblockA{\textit{School of Information and Electronics} \\
\textit{Beijing Institute of Technology}\\
Beijing, China \\
z.ding@bit.edu.cn}
\and
\IEEEauthorblockN{4\textsuperscript{th} Xiaodong~Wang}
\IEEEauthorblockA{\textit{Electrical Engineering Department} \\
\textit{Columbia University}\\
New York, USA \\
wangx@ee.columbia.edu}
}

\maketitle

\begin{abstract}
This paper concerns the problem of estimating multidimensional (MD) frequencies using prior knowledge of the signal spectral sparsity from partial time samples. In many applications, such as radar, wireless communications, and super-resolution imaging, some structural information about the signal spectrum might be known beforehand. Suppose that the frequencies lie in given intervals, the goal is to improve the frequency estimation performance by using the prior information. We study the MD Vandermonde decomposition of block Toeplitz matrices in which the frequencies are restricted to given intervals. We then propose to solve the frequency-selective atomic norm minimization by converting them into semidefinite program based on the MD Vandermonde decomposition. Numerical simulation results are presented to illustrate the good performance of the proposed method.

\end{abstract}


\section{Introduction}
Multidimensional (MD) spectral estimation from the measured signals is of practical importance in many areas of signal processing such as radar and sonar systems~\cite{skolnik1970radar,berger2010signal,nion2010tensor}, wireless communications~\cite{rappaport1996wireless,tse2005fundamentals} and super-resolution imaging~\cite{rust2006sub}. To estimate the MD spectrum, spectral estimation methods by exploiting the spectral sparsity in signals have received great attention~\cite{candes2013super,duarte2013spectral,candes2014towards,chi2015compressive,zheng2017super,li2019multi}. In particular, continuous sparse recovery methods, especially the atomic norm (AN) minimization techniques, have been proposed for spectral super-resolution~\cite{tang2013compressed,yang2014exact,bhaskar2013atomic}.

However, these compressed sensing algorithms only exploit the sparsity in spectrum and consider no prior knowledge of the measured signal. Actually, it is possible to know the information of signal spectrum a priori in many applications. For instance, in radar systems, one can set up a surveillance area where a target may appear and radar engineers may know the speed range of a particular target. The prior knowledge then enables an engineer to indicate the ballpark location of the echo from the target in the delay-Doppler spectrum. Similarly, the frequency parameters of interest can be limited in known small intervals in underwater channel estimation~\cite{beygi2015multi}. Hence, using such prior knowledge to improve the performance of super-resolution spectral estimation has attracted interest. See \cite{mishra2015spectral,chao2016extensions,yang2018frequency} and references therein. By restricting the frequency to lie in a given interval, a typical approach was proposed to solve a constrained atomic norm minimization problem for 1D frequency estimation~\cite{mishra2015spectral}. Based on the frequency-selective (FS) Vandermonde decomposition, the FS atomic norm minimization problem for 1D frequency estimation was converted into a semidefinite programming (SDP) formulation. Unfortunately, the above methods only focus on 1D frequency super-resolution problems. Besides, it's not straightforward to extend them to higher dimensional problems due to the fundamental difficulty of generalizing the classical Caratheodory's theorem~\cite{caratheodory1911zusammenhang} to higher dimensions.

In this paper, the MD-FS Vandermonde decomposition of multi-level (ML) block Toeplitz matrices for estimating MD spectrum is studied. Assume that the frequencies respectively lie in given intervals, we propose a novel method to solve the MD-FS atomic norm minimization problem by converting it into SDP formulation based on the MD-FS Vandermonde decomposition. Furthermore, we prove that the equivalence between the MD-FS atomic norm minimization and the proposed SDP formulation is  guaranteed under the condition that the ML block Toeplitz matrix is low rank. Numerical simulation results show that when prior knowledge is known, the performance advantage of the proposed method over traditional atomic norm approaches.


The remainder of the paper is organized as follows. In Section II, we set up the problem of MD spectral super-resolution with prior knowledge. In Section III, we propose the MD-FS Vandermonde decomposition results and convert the MD-FS atomic norm minimization into SDP formulation. Numerical simulation results are presented in Section IV. Section V concludes the paper.

\section{Problem Formulation}

We first introduce the 2D frequency estimation model, which is very versatile in applications, and then extend the model to MD case. Without loss of generality, consider an $N_1 \times N_2$ data matrix $\bm X_2^{\star}$, where each entry can be expressed as a superposition of $r$ complex sinusoids 
\begin{align}
x_{k_2,k_1}^{\star} = \sum_{\ell=1}^{r} \sigma_{\ell} e^{i2\pi k_1 f_{1,\ell}} e^{i2\pi k_2 f_{2,\ell}},
\end{align}
where $k_1 = 0,...,N_1-1$, $k_2 = 0,...,N_2-1$, $f_{1,\ell}, f_{2,\ell} \in {\mathbb{U}}\triangleq [0,1)$ and $\sigma_{\ell}$ are the frequencies and the complex gain associated with each $1\leq \ell \leq r$, respectively. The optimal matrix $\bm X_2^{\star}$ can be rewritten as the following matrix form
\begin{align}
\bm X_2^{\star} = \bm S_2(\bm f_2) {\rm diag}(\bm \sigma) \bm S_1^T(\bm f_1),
\end{align}
where $\bm \sigma = [\sigma_1,...,\sigma_r]^T $, ${\rm diag}(\bm \sigma)$ denotes the diagonal matrix whose diagonal entries are $\bm \sigma$ and
\begin{align}
\label{eq:S_1}
\bm S_1(\bm f_1) =&~ [\bm s_1(f_{1,1}),...,\bm s_1(f_{1,r})] \in \mathbb{C}^{N_1\times r}, \\
\label{eq:S_2}
\bm S_2(\bm f_2) =&~ [\bm s_2(f_{2,1}),...,\bm s_2(f_{2,r})] \in \mathbb{C}^{N_2\times r},
\end{align}
with $\bm s_i(f)\triangleq[1,e^{i2\pi f},...,e^{i2\pi(N_{i}-1) f}]^T\in \mathbb{C}^{N_i\times 1},~i = 1,...,d$ as the discrete complex sinusoids.

Assume that
\begin{align}
\bm Y_2 = \bm \Psi_2 \odot \bm X_2^{\star},
\end{align}
where $\odot$ is the pointwise product and $\bm \Psi_2 \in \mathbb{C}^{N_1 \times N_2}$ is the observation matrix. This model subsumes a number of signal processing systems. For example in 2D harmonic retrieval, $\bm X_2^{\star}$ is the 2D square data matrix and $\bm \Psi_2$ is a sparse sampling matrix, which observes data matrix uniformly at random~\cite{chi2015compressive}. In communication and passive radar systems, $\bm X_2^{\star}$ is the channel matrix and $\bm \Psi_2$ is the data symbol matrix~\cite{li2019multi,zheng2018adaptive}. One can vectorize $\bm Y_2$ to obtain
\begin{align}
\bm y_2 = &~{\rm vec}(\bm Y_2) =  {\rm vec}(\bm \Psi_2) \odot (\bm S_1(\bm f_1) \circ \bm S_2(\bm f_2)) \bm \sigma \nonumber \\
\label{eq:y}
=&~\bm \Phi_2 \bm x_2^{\star},
\end{align}
where $\circ$ is the Khatri-Rao product, $\bm \Phi_2 = {\rm diag}({\rm vec}(\bm \Psi_2)) \in \mathbb{C}^{N_1N_2\times N_1N_2} $ and 
\begin{align}
\bm x_2^{\star} =&~ (\bm S_1(\bm f_1) \circ \bm S_2(\bm f_2)) \bm \sigma \nonumber \\
\label{eq:xstar}
 =&~ \sum_{\ell = 1}^{r} \sigma_{\ell} \bm s_1(f_{1,\ell}) \otimes \bm s_2(f_{2,\ell}) = \sum_{\ell = 1}^{r} \sigma_{\ell}  \bm a_2(f_{1,\ell},f_{2,\ell})
\end{align}
with $\bm a_2(f_{1,\ell},f_{2,\ell})= \bm s_1(f_{1,\ell}) \otimes \bm s_2(f_{2,\ell})$ and $\otimes$ being the Kronecker product.

In the MD model, let $\bm F \in {\mathbb{U}}^{d\times r}$ be a set of $d$-dimensional frequencies $\bm f_1,...,\bm f_d \in {\mathbb{U}}^{r \times 1}$. And we define $\bm F(:,\ell) \triangleq \bm f_{:,\ell}~,\ell = 1,...,r$. Then a uniformly sampled $d$-dimensional complex sinusoid with frequency $\bm f_{:,\ell}$ and unit power can be represented by $\bm a(\bm f_{:,\ell}) = \bm a(f_{1,\ell},...,f_{d,\ell}) \triangleq \bm s_1(f_{1,\ell})  \otimes...\otimes \bm s_d(f_{d,\ell}) \in \mathbb{C}^{N_D\times 1}$ with $N_D \triangleq \prod_{i=1}^d N_i$. It follows that
\begin{align}	
\bm A \triangleq [\bm a(\bm f_{:,1}),...,\bm a(\bm f_{:,r})] = \bm S_1(\bm f_1) \circ ... \circ \bm S_d(\bm f_d) \in \mathbb{C}^{N_D\times r},
\end{align}
where $\bm S_i(\bm f_i)$ is defined with respect to $\bm s_i(f)$ similarly as in \eqref{eq:S_1} and \eqref{eq:S_2}. In the problem of MD frequency estimation, the vectorized data $\bm y \in \mathbb{C}^{N_D\times 1}$ follows a similar parametric model:
\begin{align}
\label{eq:y-1}
\bm y = \bm \Phi \bm x^{\star} = \bm \Phi  \sum_{\ell = 1}^{r} \sigma_{\ell}  \bm a(\bm f_{:,\ell}) .
\end{align}
In this paper, we consider the case that frequencies are known a priori as $\bm f_i \in {\mathbb{F}}_{i}^{r \times 1} = [f_{L,i},f_{H,i}]^{r \times 1} \in {\mathbb{U}}^{r \times 1},~i=1,...,d$, where ${\mathbb{F}} = [f_L,f_H]$ denotes a closed interval as usual if $f_L< f_H$. Otherwise, we define ${\mathbb{F}} = [f_L,f_H]\triangleq {\mathbb{U}} \backslash (f_H,f_L)$ if $f_L> f_H$. Our aim is to estimate $\bm F$ in $\bm x^{\star}$ under the prior constraints from the observation $\bm y$ in \eqref{eq:y-1}.

\section{Proposed Method Based on FS Atomic norm}

In this section, we present the proposed method for the MD spectral super-resolution based on the MD-FS atomic norm. First the MD-FS atomic norm is used to setup the optimization problem for MD spectral super-resolution. Then the MD-FS Vandermonde decomposition result of ML block Toeplitz matrices is presented, which enables us to convert the MD-FS atomic norm optimization problem into SDP formulation.

\subsection{Setup the Problem Based on MD-FS Atomic Norm}


We define the MD-FS atomic set as the collection of all MD complex sinusoids:
\begin{align}
{\cal A}({\mathbb{F}})\triangleq\{  \bm a(f_1,...,f_d): f_i \in {\mathbb{F}}_{i},~i=1,...,d \},
\end{align}
then the MD-FS atomic norm with respect to signal $\bm x^{\star}$ in \eqref{eq:y-1} is given in the following definition.
\begin{definition}
	The MD-FS atomic norm for $\bm x$ in \eqref{eq:y-1} is 
	\begin{align}
\label{eq:2DAN-definition}
\|\bm x\|_{{\cal A}({\mathbb{F}})}\triangleq &~ \inf \{ \chi>0: \bm x \in \chi {\rm conv}({\cal A}({\mathbb{F}})) \} \nonumber \\
=&~\inf_{\substack{f_{i,\ell} \in {\mathbb{F}}_{i}, i=1,...,d \\ \sigma_{\ell} \in \mathbb{C}}} \left\{     \sum_{\ell} |\sigma_{\ell}| : \bm x =  \sum_{\ell} \sigma_{\ell}  \bm a(\bm f_{:,\ell})  \right\}.
\end{align}
\end{definition}

By introducing the spectral sparsity, our MD frequency estimation problem can be formulated according to \eqref{eq:y-1} as
\begin{align}
\label{eq:AN-problem-1}
\bm{\widehat{x}} = \arg \min_{\bm x \in \mathbb{C}^{N_D \times 1} } \| \bm x\|_{{\cal A}({\mathbb{F}})},~{\text{s.t.}}~  \bm y = \bm \Phi \bm x.
\end{align}
This shows that we find the optimal $\bm x$ by seeking a signal with minimum MD-FS atomic norm satisfying the observation constraints. After $\bm x$ is obtained from \eqref{eq:AN-problem-1}, the frequencies and complex gains $\bm \sigma$ in $\bm x$ can be determined by using the MD MUltiple SIgnal Classifier (MD-MUSIC)~\cite{swindlehurst1992performance} algorithm with $\bm x$ as an input. In particular, the MD-MUSIC method determines the frequencies by locating the poles in the spectrum and estimates the complex gains by using least square  method with the estimated frequencies.

Note that the MD-FS atomic norm in \eqref{eq:AN-problem-1} is essentially semi-infinite programming, it cannot be directly solved. We will show  how to solve \eqref{eq:AN-problem-1} based on the MD-FS Vandermonde decomposition in the following subsections.

\subsection{MD-FS Vandermonde Decomposition of ML Block Toeplitz Matrices}

Note that for any $\ell$, an atom in the form 
\begin{equation}
\bm a(\bm f_{:,\ell}) \bm a^H(\bm f_{:,\ell})  = (\bm s_1(f_{1,\ell}) \bm s_1^H(f_{1,\ell}))  \otimes...\otimes (\bm s_d(f_{d,\ell}) \bm s_d^H(f_{d,\ell}))
\end{equation}
forms a $d$-level block Toeplitz matrix $\bm T^{d} \triangleq \bm T^{d}({\cal B}^d) \in \mathbb{C}^{N_D \times N_D}$. In particular, for a $d$-way tensor ${\cal B}^d \in \mathbb{C}^{(2N_1-1)\times...\times(2N_d-1)}$, $\bm T^{d}({\cal B}^d)$ is defined as taking ${\cal B}^d$ as an input and outputing recursively as
\begin{align}
\label{eq:level-Toeplitz}
&{\bm T^{d}}({\cal B}^d) = \nonumber \\
&\left[ {\begin{array}{*{20}{c}}
	{\bm T^{d-1}}({\cal B}^{d-1}(0)) &  \cdots & {\bm T^{d-1}}({\cal B}^{d-1}(N_d-1))\\
	{\bm T^{d-1}}({\cal B}^{d-1}(-1)) &  \cdots & {\bm T^{d-1}}({\cal B}^{d-1}(N_d-2))\\
	\vdots &  \ddots & \vdots \\
	{\bm T^{d-1}}({\cal B}^{d-1}(1-N_d)) &  \cdots & {\bm T^{d-1}}({\cal B}^{d-1}(0))
	\end{array}} \right],
\end{align}
where ${\cal B}^{d-1}(i) = {\cal B}^{d}(:,...,:,i)$. For $d=1$ we have \eqref{eq:level-Toeplitz} reduces to the standard Topelitz matrix as
\begin{align}
\label{eq:Toeplitz}
{{\rm Toep}(\bm b)} =
\left[ {\begin{array}{*{20}{c}}
	{b_{0}} & {b_{-1}} & \cdots & {b_{-N_1+1}}\\
	{b_{1}} & {b_{0}} & \cdots & {b_{-N_1+2}}\\
	\vdots & \vdots & \ddots & \vdots \\
	{b_{N_1-1}} & {b_{N_1-2}} & \cdots & {b_{0}}
	\end{array}} \right] \in \mathbb{C}^{N_1 \times N_1},
\end{align}
where $b_j$ denotes the $j$-th element in $\bm b \in \mathbb{C}^{2 N_1-1}$.
From \eqref{eq:level-Toeplitz}, we can have
\begin{align}
&{\bm T^{d}}(m_1,n_1;m_2,n_2;...;m_d,n_d)  \nonumber \\
&~~~~~~~~~~~~~~~= {\cal B}^d(m_1-n_1,m_2-n_2,...,m_d-n_d),\\
&~~~~~~~~~~~~~~m_1,n_1 = 1,...,N_1;...;m_d,n_d=1,...,N_d, \nonumber
\end{align}
where ${\bm T^{d}}(...;m_i,n_i;...)$ denotes the $(m_i,n_i)$-th element or block in the $i$-th level of $\bm T^{d}$.

To solve an atomic norm minimization problem, the idea is to convert it into a semidefinite program based on the Vandermonde decomposition of Toeplitz matrix. We therefore present the MD-FS Vandermonde decomposition result of ML block Toeplitz matrices in this subsection for solving \eqref{eq:AN-problem-1}. To begin with, we first introduce the following theorem.

\begin{theorem}(Theorem 1, \cite{yang2016vandermonde2}) \label{the: MDVD} Assume that $\bm T^d$ is a PSD $d$-level block Toeplitz matrix with $d \geq 1$ and $r = {\rm rank}(\bm T^d) < \min_i N_i$. Then, $\bm T^d$ can be decomposed as
\begin{align}
\label{eq:VSV}
\bm{T}^d = \bm A \bm \Sigma \bm A^H = \sum_{\ell=1}^{r} \sigma_{\ell} \bm{a}\left(\bm{f}_{:,\ell}\right) \bm{a}^{H}\left(\bm{f}_{:,\ell}\right),
\end{align}
where $\bm \Sigma = {\rm diag}([\sigma_1,...,\sigma_r]) \in \mathbb{C}^{r \times r}$ with $\sigma_{\ell}>0,~\bm f_{:,\ell},~\ell = 1,...,r$ are distinct points in ${\mathbb{U}}^{d\times 1}$, and the $(d + 1)$-tuples $(\bm f_{:,\ell},\sigma_{\ell}),~\ell = 1,...,r$ are unique.
\end{theorem}

    The above theorem shows that once $r = {\rm rank}(\bm T^d) < \min_i N_i$ holds, the $d$-level block Toeplitz matrix ${\bm T}^d$ has the MD Vandermonde decomposition in \eqref{eq:VSV} if ${\bm T}^d \succeq 0$. In order to combine the interval information into MD Vandermonde decomposition, we show the property of trigonometric polynomials in the following lemma, which is proved in~\cite{li2018multidimensional}.

\begin{lemma} \label{lem: ptp} If
\begin{align}
\label{eq:r0}
r_{0,i} =&~ -2 \cos[\pi(f_{H,i}-f_{L,i})] {\rm sign}(f_{H,i}-f_{L,i}), \\
\label{eq:r1}
r_{1,i} =&~ e^{i\pi(f_{L,i}+f_{H,i})} {\rm sign}(f_{H,i}-f_{L,i}),
\end{align}
where ${\rm sign}(\cdot)$ denotes the sign function. Then, when $f_{L,i} \neq f_{H,i}$, the trigonometric polynomials 
\begin{align}
\label{eq:g1}
g_i(f_i) =&~ r_{1,i} x_i^{-1} + r_{0,i} + r_{-1,i} x_i \nonumber \\
=&~ r_{0,i} + 2\Re\{r_{1,i}e^{-i2\pi f_i}\},
\end{align}
are always positive on $(f_{L,i},f_{H,i})$ and negative on $(f_{H,i},f_{L,i})$ for $i=1,...,d$, where $\Re$ returns the real part of a complex argument.
\end{lemma}

The above lemma shows that we can restrict the frequencies in given intervals respectively by setting $g_i(f_i)\geq 0,~i=1,...,d$. To this end, we introduce the MD-FS Vandermonde decomposition of $d$-level block Toeplitz matrices in the following theorem.

\begin{theorem} \label{thm: FSVD} For a $d$-level block Toeplitz matrix $\bm T^d \in \mathbb{C}^{N_D\times N_D}$ with $d\geq 2$, if $r = {\rm rank}(\bm T^d) < \min_i N_i$, then given ${\mathbb{F}}_{i} \in {\mathbb{U}},~i=1,...,d$, it has an MD-FS Vandermonde decomposition as in \eqref{eq:VSV} with $f_{i,\ell} \in {\mathbb{F}}_{i},~\ell = 1,...,r,~i=1,...,d$, if and only if 
\begin{align}
\label{eq:T}
\bm T^d &\succeq \bm 0, \\
\label{eq:Tp}
\bm T_{g_i}^d &\succeq \bm 0,~i=1,...,d,
\end{align}
where $g_i$ is defined by \eqref{eq:g1} and the $d$-level block Toeplitz matrices $\bm T^d_{g_i} \in \mathbb{C}^{N_{D-1} \times N_{D-1}},~i=1,...,d$ with $N_{D-1} \triangleq \prod_{i=1}^d (N_i-1)$ are defined respectively for $i=1,...,d$ as 
\begin{align}
\label{dTgd}
&{\bm T^{d}_{g_i}}(m_1,n_1;m_2,n_2;...;m_d,n_d) \nonumber \\
=&~ \sum_{k=-1}^{1} r_{k,i} {\cal B}^d(m_1-n_1,...,m_i-n_i-k,...,m_d-n_d), \nonumber \\
&~m_1,n_1 = 1,...,N_1-1;...;m_d,n_d=1,...,N_d-1. 
\end{align}
\end{theorem}

\begin{IEEEproof}
We first prove the sufficient condition. Following from \eqref{eq:T} and Theorem~\ref{the: MDVD}, we know that $\bm T^d$ has an MD Vandermonde decomposition as in \eqref{eq:VSV}. Therefore, we need to prove $f_{i,\ell} \in {\mathbb{F}}_{i},~\ell = 1,...,r,~i=1,...,d$ under the additional conditions \eqref{eq:Tp}. For the MD Vandermonde decomposition in \eqref{eq:VSV}, we have
\begin{align}
&{\cal B}^d(m_1-n_1,...,m_d-n_d) = \bm T^d(m_1,n_1;...;m_d,n_d) \nonumber \\
&= \sum_{\ell=1}^r \sigma_{\ell} e^{i2\pi(m_1-n_1)f_{1,\ell}} \times...\times e^{i2\pi(m_d-n_d)f_{d,\ell}},
\end{align}
which shows that for $i = 1,...,d$
\begin{align}
&\bm T_{g_i}^d(m_1,n_1;...;m_d,n_d) \nonumber \\
=&~ \sum_{k=-1}^{1} r_{k,i} {\cal B}^d(m_1-n_1,...,m_i-n_i-k,...,m_d-n_d)  \nonumber \\
=&~ \sum_{k=-1}^{1} r_{k,i} \sum_{\ell=1}^r \sigma_{\ell} e^{i2\pi(m_1-n_1)f_{1,\ell}} \times \nonumber \\
&~ ... \times e^{i2\pi(m_i-n_i-k)f_{i,\ell}} \times ... \times e^{i2\pi(m_d-n_d)f_{d,\ell}} \nonumber \\
=&~ \sum_{k=-1}^{1} r_{k,i} e^{-i2\pi k f_{i,\ell}} \sum_{\ell=1}^r \sigma_{\ell} e^{i2\pi(m_1-n_1)f_{1,\ell}} \times \nonumber \\
&~...\times e^{i2\pi(m_d-n_d)f_{d,\ell}}  \nonumber \\
=&~ \sum_{\ell=1}^r \sigma_{\ell} g_i(f_{i,\ell}) e^{i2\pi(m_1-n_1)f_{1,\ell}} \times...\times e^{i2\pi(m_d-n_d)f_{d,\ell}},\nonumber \\
&~m_1,n_1 = 1,...,N_1-1;...;m_d,n_d=1,...,N_d-1,
\end{align}
implies that,
\begin{align}
\label{eq:AdA}
\bm T_{g_i}^d = \bm{\bar{A}} {\rm diag}\left([\sigma_1 g_i(f_{i,1}),...,\sigma_{r} g_i(f_{i,r})]^T \right) \bm{\bar{A}}^H,
\end{align}
where 
\begin{align}	
\bm{\bar{A}} &\triangleq [\bm{\bar{a}}(\bm f_{:,1}),...,\bm{\bar{a}}(\bm f_{:,r})] \in \mathbb{C}^{N_{D-1}\times r}, \\
\bm{\bar{a}}(\bm f_{:,\ell}) &\triangleq \bm{\bar{s}}_1(f_{1,\ell})  \otimes...\otimes \bm{\bar{s}}_d(f_{d,\ell}) \in \mathbb{C}^{N_{D-1}\times 1}, \\
&~~~~~~~~~~~~~~~~~~~~~~~~~~~~~~~~~~\ell = 1,...,r, \nonumber \\
\bm{\bar{s}}_i(f)  &\triangleq[1,e^{i2\pi f},...,e^{i2\pi(N_{i}-2) f}]^T\in \mathbb{C}^{(N_i-1)\times 1}, \\
&~~~~~~~~~~~~~~~~~~~~~~~~~~~~~~~~~~i = 1,...,d.\nonumber
\end{align}
Since $r < \min_i N_i$, $\bm{\bar{A}}$ has full column rank. Then, by noting \eqref{eq:AdA} and \eqref{eq:Tp} we have for $i=1,...,d$
\begin{align}
\label{eq:ATpA}
{\rm diag}\left([\sigma_1 g_i(f_{i,1}),...,\sigma_{r} g_i(f_{i,r})]^T \right) = \bm{\bar{A}}^{\dagger} \bm T_{g_i}^d \bm{\bar{A}}^{\dagger H} \geq 0,
\end{align}
where $(\cdot)^{\dagger}$ denotes the matrix pseudo-inverse operator. \eqref{eq:ATpA} implies that $\sigma_{\ell} g_i(f_{i,\ell})\geq 0,~i=1,...,d$, which immediately follows that $g_i(f_{i,\ell})\geq 0,~i=1,...,d$ since $\sigma_{\ell}>0,~\ell = 1,...,r$. By noting Lemma~\ref{lem: ptp} we finally have $f_{i,\ell} \in {\mathbb{F}}_{i},~\ell = 1,...,r$.

Next we prove the necessary condition. Given $\bm{T}^d$ in \eqref{eq:VSV} with $f_{i,\ell} \in {\mathbb{F}}_{i},~\ell = 1,...,r,~i=1,...,d$. We have \eqref{eq:T} holds since $\sigma_{\ell}>0,~\ell = 1,...,r$. Moreover, \eqref{eq:Tp} also holds since we have $g_i(f_{i,\ell})\geq 0,~i=1,...,d,~\ell = 1,...,r$ in \eqref{eq:ATpA} by noting Lemma~\ref{lem: ptp}. Therefore we complete the proof.
\end{IEEEproof}

It is noteworthy that the MD-FS Vandermonde decomposition result can be extended to the multiple frequency band case, which is given in the following corollary. The proof can be found in~\cite{li2018multidimensional}. In the next subsection, we convert \eqref{eq:AN-problem-1} into convex SDP formulation based on the MD-FS Vandermonde decomposition.

\begin{corollary} \label{cor: FSVD} For a $d$-level block Toeplitz matrix $\bm T^d({\cal B}^d) \in \mathbb{C}^{N_D\times N_D}$, if $r = {\rm rank}(\bm T^d({\cal B}^d)) < \min_i N_i$, it admits an MD-FS Vandermonde decomposition as in \eqref{eq:VSV} with $f_{i,\ell} \in \bigcup_{j}^J {\mathbb{F}}_{i,j},~\ell = 1,...,r,~i=1,...,d$ where ${\mathbb{F}}_{i,j} = [f_{L,i,j},f_{H,i,j}] \in {\mathbb{U}},~j=1,...,J,~i=1,...,d$, if and only if there exist $d$-way tensors ${\cal B}^d_j$ satisfying
\begin{align}
\label{eq:cor-B}
&~\sum_{j=1}^J {\cal B}^d_j = {\cal B}^d, \\
\label{eq:cor-rank}
&~\sum_{j=1}^J {\rm rank}(\bm T^d({\cal B}^d_j)) = r, \\
\label{eq:cor-T-c}
&~\bm T^d({\cal B}^d_j) \succeq  \bm 0, \\
\label{eq:cor-Tp-c}
&~\bm T_{g_{i,j}}^d({\cal B}^d_j) \succeq  \bm 0,~i = 1,...,d
\end{align}
for $j = 1,...,J$, where $g_{i,j}$ is defined with respect to $[f_{L,i,j},f_{H,i,j}]$.
\end{corollary}

\subsection{SDP Formulation of MD-FS Atomic Norm}

Under the condition ${\rm rank}(\bm T^d) < \min_i N_i $, the MD-FS atomic norm minimization can be converted to SDP formulation by applying the MD-FS Vandermonde decomposition in the following theorem.

\begin{theorem} \label{prop: FSAN} For the MD-FS atomic norm defined in~\eqref{eq:2DAN-definition}, we have that
\begin{align}
\label{eq:FSAN-thm}
\|\bm x\|_{{\cal A}({\mathbb{F}})} \geq &~ \min_{{\cal B}^d, {t}}  \frac{1}{2N_D} {\rm Tr}(\bm T^d({\cal B}^d)) + \frac{1}{2} {t}, \\
&~{\text{s.t.}}~\left[ {\begin{array}{*{20}{c}}
	\bm T^d({\cal B}^d) & \bm x \\
	\bm x^H & {t}
	\end{array}} \right] \succeq 0,
	\nonumber \\
&~~~~~~\bm T_{g_i}^d({\cal B}^d) \succeq 0,~i=1,...,d, \nonumber
\end{align}
where $g_i$ and $\bm T_{g_i}^d$ are defined by \eqref{eq:g1} and \eqref{dTgd}, respectively. And if ${\rm rank}(\bm T^d({\cal B}^d)) < \min_i N_i$, we further have $\|\bm x\|_{{\cal A}({\mathbb{F}})}$ equals to the right-hand side of \eqref{eq:FSAN-thm}.
\end{theorem}

\begin{IEEEproof} Denote the value of the right-hand side of \eqref{eq:FSAN-thm} by ${\rm SDP}{(\bm x)}$. Let $\bm x = \sum_{\ell} \sigma_{\ell} \bm a(\bm f_{:,\ell})$ be an MD-FS atomic decomposition of $\bm x$ on ${\mathbb{F}}_{i},~i=1,...,d$, where $\sigma_{\ell} = |\sigma_{\ell}| e^{ i \theta_{\ell}}$. Then we have
\begin{align}
\sum_{\ell} |\sigma_{\ell}| \left[ {\begin{array}{*{20}{c}}
	{\bm a(\bm f_{:,\ell})}\\
	{e^{i\theta_{\ell}}}
	\end{array}} \right] 
	\left[ {\begin{array}{*{20}{c}}
	{\bm a(\bm f_{:,\ell})}\\
	{e^{i\theta_{\ell}}}
	\end{array}} \right]^H = 
 \left[ {\begin{array}{*{20}{c}}
	\bm T^d & \bm x\\
	\bm x^H& {\sum\limits_{\ell} |\sigma_{\ell}|}
	\end{array}} \right] \succeq 0,
\end{align}
where $\bm T^d = \sum_{\ell} |\sigma_{\ell}|{\bm a(\bm f_{:,\ell})}{\bm a(\bm f_{:,\ell})}^H$. Hence, we also have $\bm T_{g_i}^d \succeq 0,~i=1,...,d$ by noting \eqref{eq:ATpA} and Lemma~\ref{lem: ptp}. Now,
\begin{align}
\label{eq:SDP-inequality-1}
{\rm SDP}{(\bm x)}&\leq \frac{1}{2N_D}{\rm{Tr}}(\bm T^d) + \frac{1}{2}  {\sum_{\ell} |\sigma_{\ell}|}\nonumber \\
&= {\sum_{\ell} |\sigma_{\ell}|} = \|\bm x\|_{{\cal A}({\mathbb{F}})}.
\end{align}
Hence, we can have \eqref{eq:FSAN-thm} holds.

On the other hand, if ${\rm rank}(\bm T^d({\cal B}^d)) < \min_i N_i$ holds. Suppose the optimal ${\widehat{\cal B}}^d$ and $\widehat {t}$ of the problem in \eqref{eq:FSAN-thm} satisfy
\begin{eqnarray}
\left[ {\begin{array}{*{20}{c}}
	\bm T^d({\widehat{\cal B}}^d) & \bm x\\
	{\bm x^H}& {{t}}
	\end{array}} \right] \succeq 0,~\bm T_{g_i}^d({\widehat{\cal B}}^d) \succeq 0,~i=1,...,d,
\end{eqnarray}
then we have $\bm T^d({\widehat{\cal B}}^d) \succeq 0$ and $\bm T^d({\widehat{\cal B}}^d) \succeq t^{-1}\bm x \bm x^H$ by the Schur complement condition. It immediately follows that $\bm T^d({\widehat{\cal B}}^d)$ has an MD-FS Vandermonde decomposition as in \eqref{eq:VSV} with $f_{i,\ell} \in {\mathbb{F}}_{i},~\ell = 1,...,r,~i=1,...,d$ according to Theorem~\ref{thm: FSVD}. Then, we have that $\bm x$ lies in the column space of $\bm T^d({\widehat{\cal B}}^d)$ and $\bm x = \bm A \bm \sigma$ for some vector $\bm \sigma$. Bring $\bm x = \bm A \bm \sigma$ into $\bm T^d({\widehat{\cal B}}^d) \succeq t^{-1}\bm x \bm x^H$ yields 
\begin{align}
\bm A \bm \Sigma \bm A^H \succeq {{t}}^{-1} \bm A \bm \sigma \bm \sigma^H \bm A^H. 
\end{align}
Let $\bm z$ be any vector such that $\bm A^H \bm z = {\rm sign}(\bm \sigma)$. Then the following inequality holds
\begin{align}
{\rm Tr}(\bm \Sigma)  &= \bm z^H \bm A \bm \Sigma \bm A^H \bm z \geq {{t}}^{-1}\bm z^H \bm A \bm \sigma \bm \sigma^H \bm A^H \bm z \nonumber \\
&= {{t}}^{-1}(\sum_{\ell}|\sigma_{\ell}|)^2.
\end{align}
Hence
\begin{align}
\frac{1}{2N_D}{\rm{Tr}}(\bm T^d({\widehat{\cal B}}^d)) + \frac{1}{2} \widehat {t} =&~\frac{1}{2} {\rm{Tr}}(\bm \Sigma) + \frac{1}{2} \widehat {t}  \geq \sqrt{{\rm{Tr}}(\bm \Sigma) \widehat {t} } \nonumber \\ 
\geq&~  \sum_{\ell}|\sigma_{\ell}| \geq \|\bm x\|_{{\cal A}({\mathbb{F}})},
\end{align}
which is equivalent to ${\rm SDP}{(\bm x)} \geq \|\bm x\|_{{\cal A}({\mathbb{F}})}$. Combining the inequality with \eqref{eq:SDP-inequality-1} we conclude $\|\bm x\|_{{\cal A}({\mathbb{F}})} = {\rm SDP}{(\bm x)}$ if ${\rm rank}(\bm T^d) < \min_i N_i$, which completes the proof.
\end{IEEEproof}


By applying Theorem~\ref{prop: FSAN} we can approximately{\footnote{Although the SDP is an approximation, the simulation results show that the performance is good even if the condition ${\rm rank}(\bm T^d) < \min_i N_i$ is not satisfied.}} convert \eqref{eq:AN-problem-1} into the following SDP:
\begin{align}
\label{eq:SDP-problem-1}
& \min_{\bm x, {\cal B}^d, {t} } \frac{1}{2N_D} {\rm Tr}(\bm T^d({\cal B}^d)) + \frac{1}{2} {t}, \\
&~{\text{s.t.}}~\bm y = \bm \Phi \bm x,~\left[ {\begin{array}{*{20}{c}}
	\bm T^d({\cal B}^d) & \bm x \\
	\bm x^H & {t}
	\end{array}} \right] \succeq 0, \nonumber \\
&~~~~~\bm T_{g_i}^d({\cal B}^d) \succeq 0,~i=1,...,d. \nonumber
\end{align}
After $\bm x$ is obtained from \eqref{eq:SDP-problem-1}, as mentioned in Section III-A, the frequencies can be determined by the MD-MUSIC algorithm. Optimization problem \eqref{eq:SDP-problem-1} is convex, hence it can be solved with standard convex solvers, e.g., CVX~\cite{boyd2004convex}. Assume that the number of the positive semidefinite constraints in \eqref{eq:SDP-problem-1} is $N_p$, then the complexity in each iteration is ${\cal O}(N_pN_D^6)$ if the interior point method is used.

\section{Numerical Simulations}

Since the 2D frequency estimation problem is very common in practice, i.e., in high-resolution radar systems and wireless communications, there frequency pairs correspond to delay, Doppler and magnitudes of the scatterers. We hence present numerical examples in this section for a data matrix $\bm X^{\star}$ of size $N_1\times N_2$ with $N_1 = N_2 = 8$.
In the simulations, the coefficient of each frequency is generated with fixed magnitude one and random phase, and frequency pairs are randomly generated in $[0.3,0.4)\times[0.5,0.6)$. Hence, the prior knowledge is set as $f_{L,1} = 0.3$, $f_{H,1} = 0.4$, $f_{L,2} = 0.5$ and $f_{H,2} = 0.6$. In addition, $\bm \Phi$ is set as a diagonal matrix with some elements on the diagonal are equal to 1 and the other elements are equal to 0 (corresponding to the case that $\bm \Psi$ is the sampling matrix).

The traditional 2D AN method in \cite{chi2015compressive} is used as the baseline for comparison, which can be converted into similar convex SDP problem as in \eqref{eq:SDP-problem-1} but without the further constraints $\bm T_{g_i}^d({\cal B}^d) \succeq 0,~i=1,...,d$.

\begin{figure}[!htb]
	\centering
		
	\subfloat[][]{\includegraphics[width=2.3in]{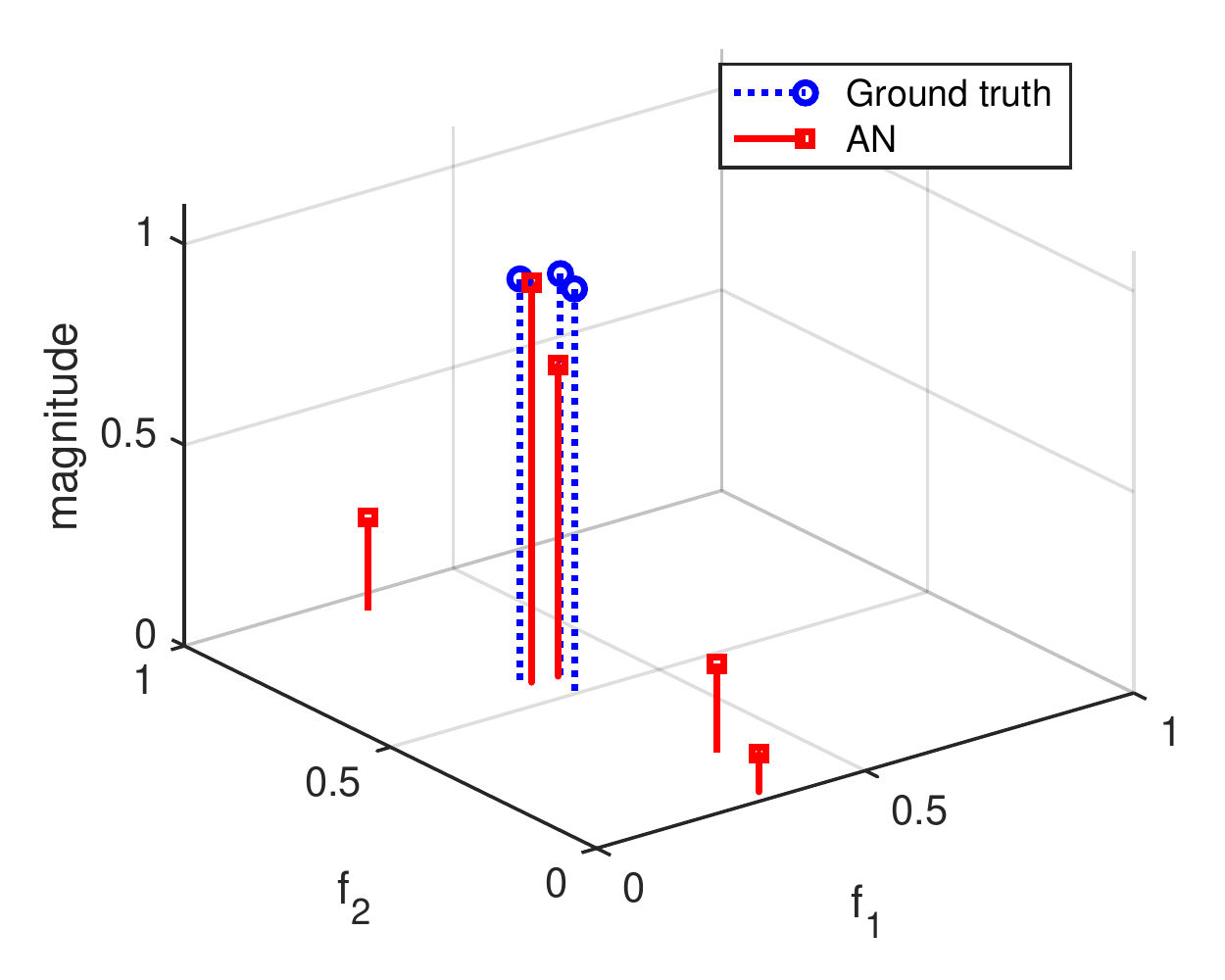}}
	
	\subfloat[][]{\includegraphics[width=2.3in]{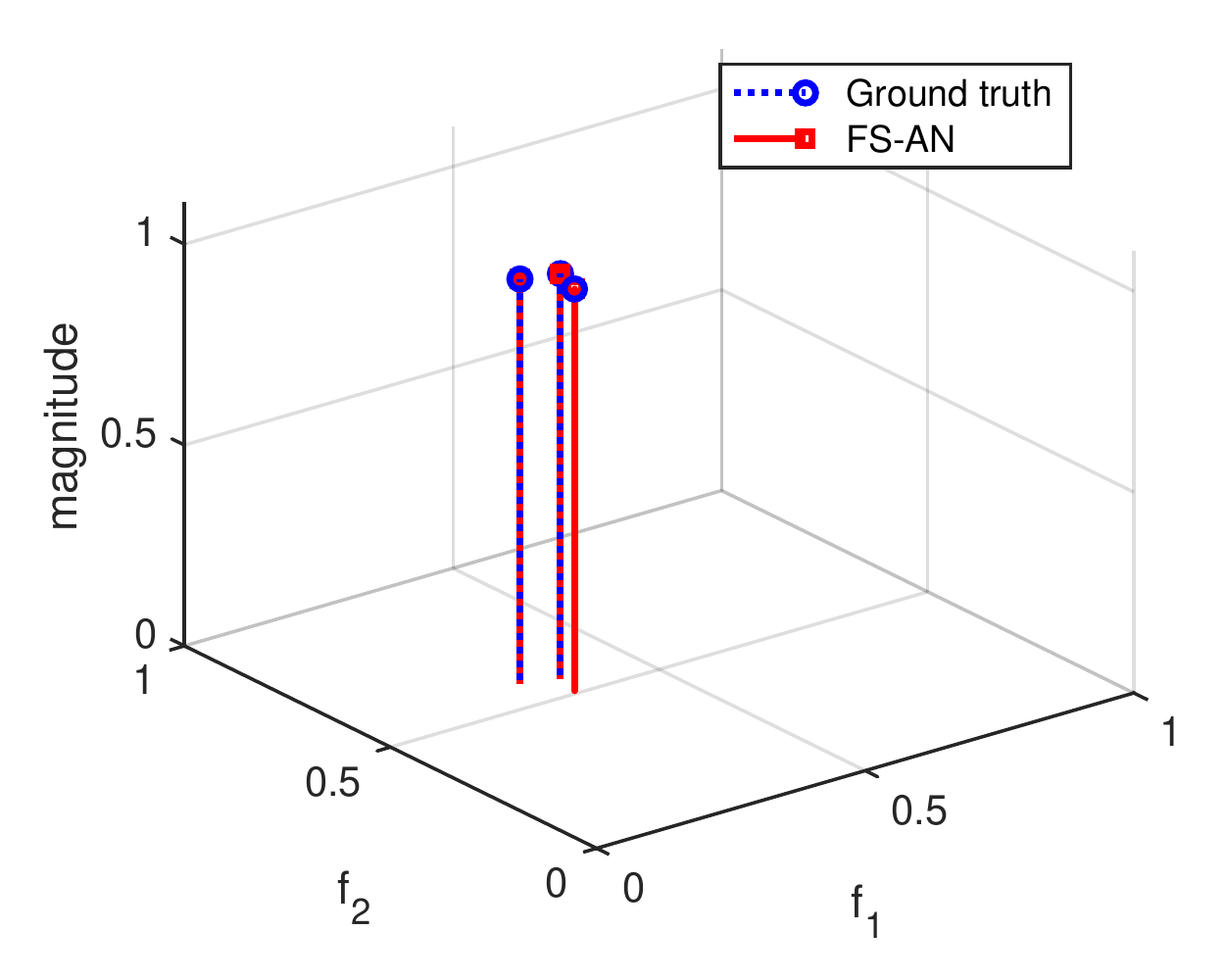}}
	
	\caption{Frequency estimation results. (a) AN; (b) FS-AN.}
	\label{figure:frequency-results}
\end{figure}

We first present an example when $r=3$ with $\bm f_1 = [0.35,0.31,0.37]^T$ and $\bm f_2 = [0.51,0.59,0.57]^T$ to demonstrate the effectiveness of the proposed method on frequency estimation. The number of samples (number of non-zero elements in $\bm \Psi$) is set as $N_s = 12$. In Fig.~\ref{figure:frequency-results}, the frequency estimation results of the AN and FS-AN methods are presented. The 2D-MUSIC~\cite{zheng2017super,berger2010signal} is used to localize the frequencies after $\bm{\widehat{x}}$ is available. We can see that with the prior knowledge of frequency ranges, the FS-AN method still work well under high $r/N_s$, while the traditional AN method suffers dramatic degradation. 

\begin{figure*}[!htb]
	\centering
		
	\subfloat[][]{\includegraphics[width=2.3in]{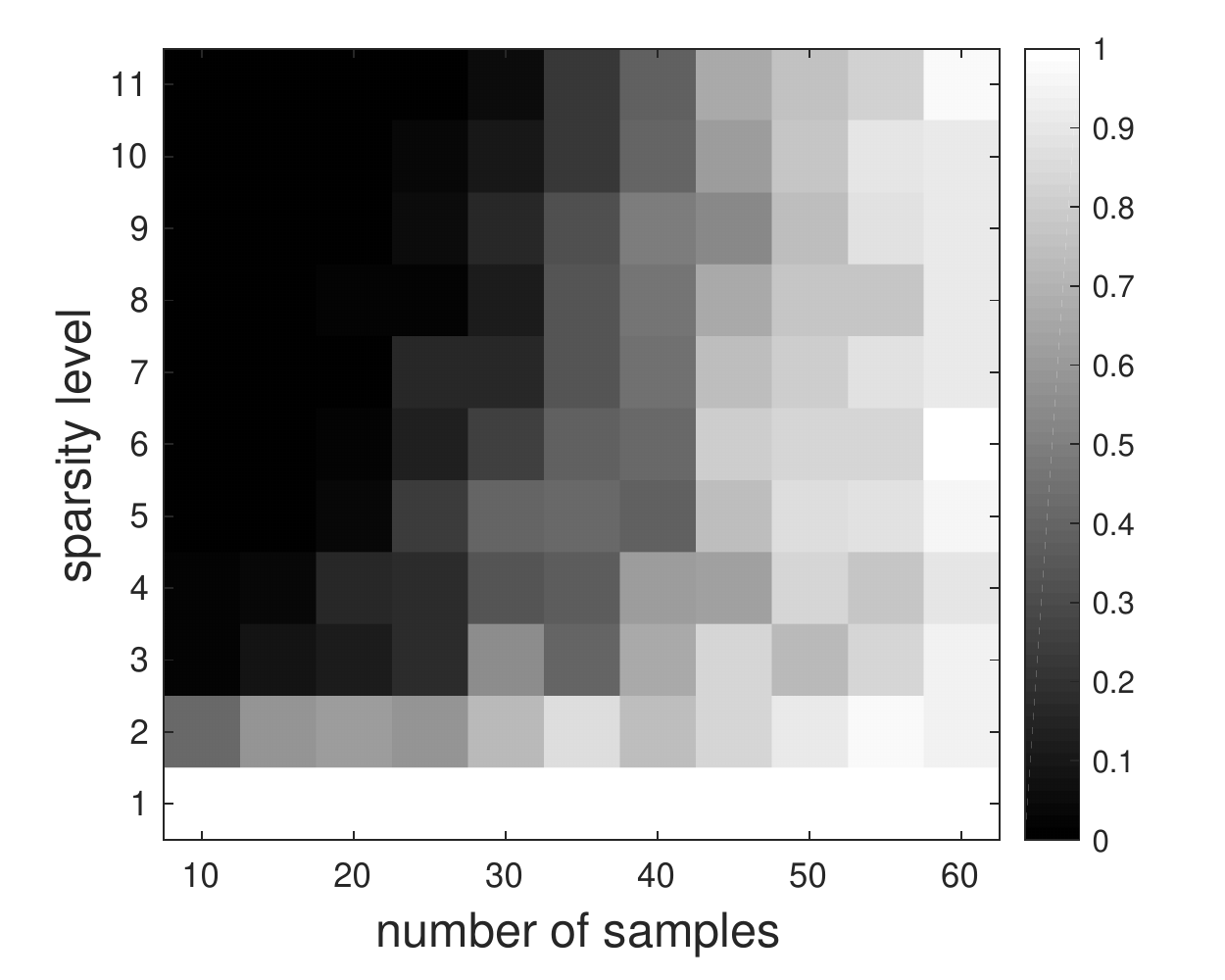}}
	\subfloat[][]{\includegraphics[width=2.3in]{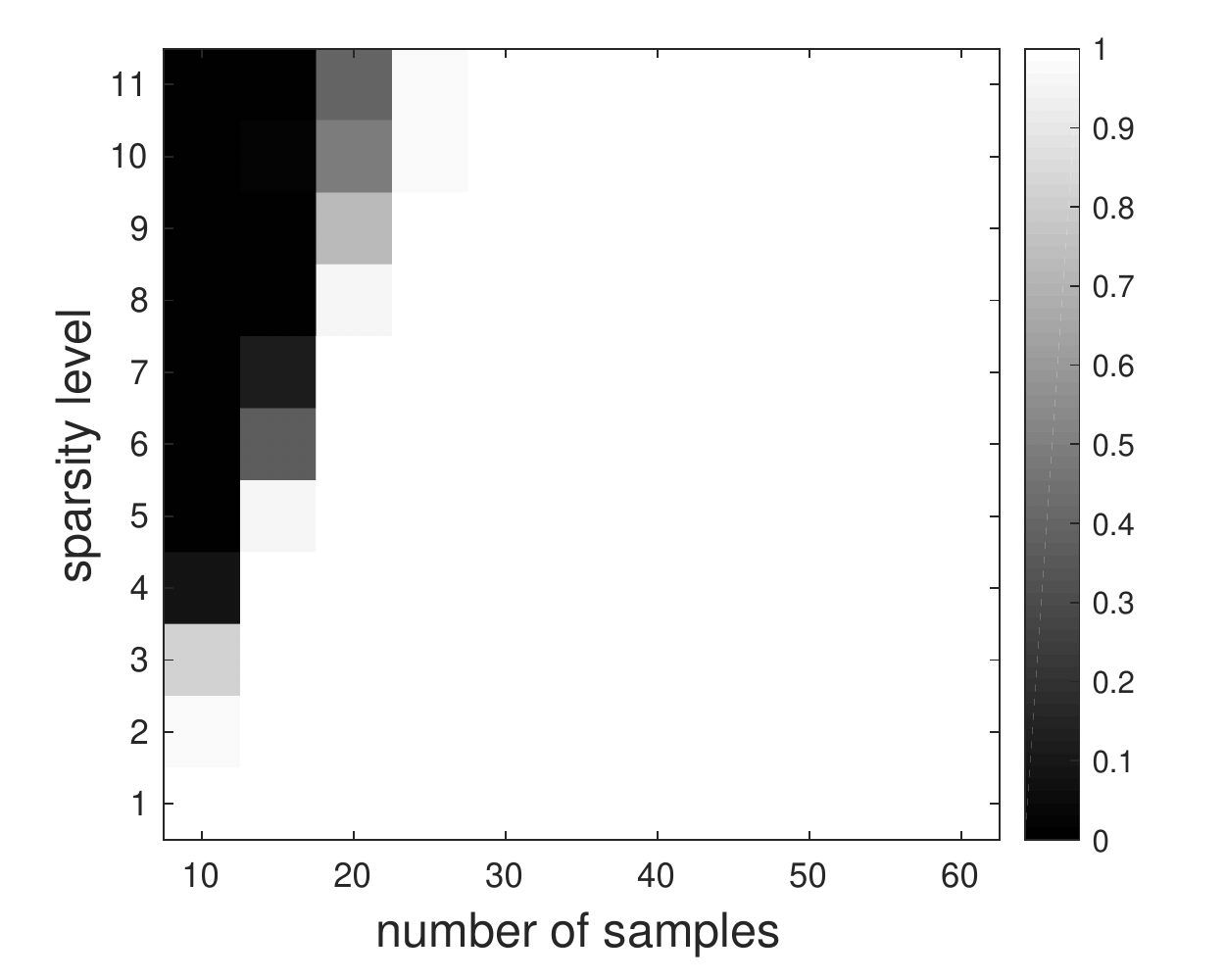}}
	\subfloat[][]{\includegraphics[width=2.3in]{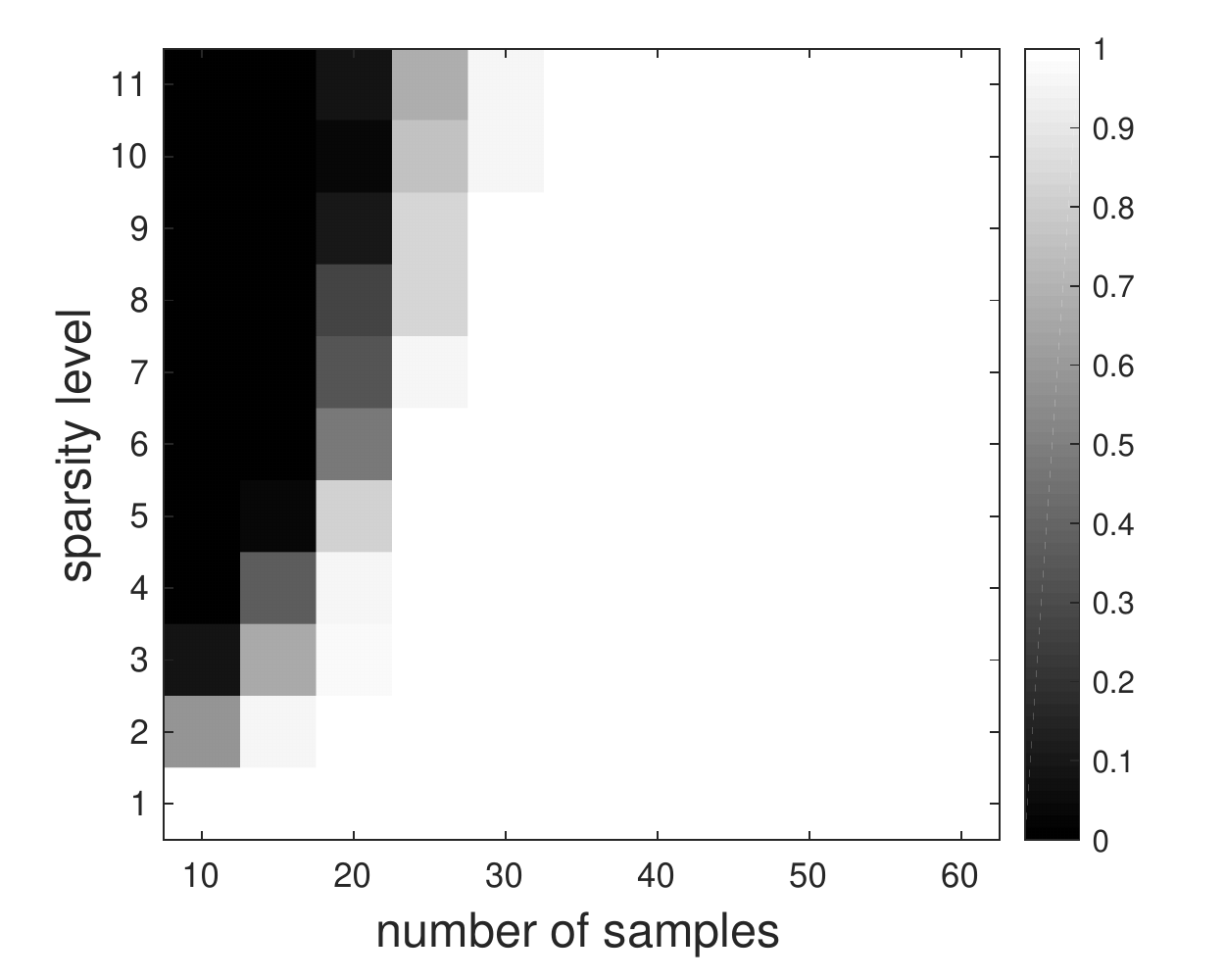}}
	
	\caption{Phase transition plots when $N_1 = N_2 = 8$. (a) AN; (b) FS-AN; (c) FS-AN with rough prior knowledge. The success rate is calculated by averaging over 50 runs. The grayscale of each cell reflects the empirical rate of success.}
	\label{figure:phase-transition}
\end{figure*}

We next examine the phase transition of the proposed method. For each number of samples and sparsity level $r$, we run 10 experiments. In each experiments, the recovery was claimed successful if the normalized mean-squared-error (NMSE) $\|\bm{\widehat x} - \bm x^{\star}\|_2 / \|\bm x^{\star}\|_2$ is smaller than $10^{-5}$. Fig.~\ref{figure:phase-transition} shows the success rate for each $N_s$ and $r$, with the grayscale of each cell reflecting the empirical rate of success. Note that sometimes people may only know rough ranges of the frequencies, hence we also evaluate the proposed method with rough prior knowledge as $f_{L,1} = 0.2$, $f_{H,1} = 0.4$, $f_{L,2} = 0.5$ and $f_{H,2} = 0.7$. Comparing Fig.~\ref{figure:phase-transition}(b) with (a), it can be seen that when the accurate prior knowledge of frequency ranges is known, the performance of the 2D harmonic retrieval can be significantly improved. And from Fig.~\ref{figure:phase-transition}(c) we can see that when the prior knowledge is not very accurate, but contains the real frequency ranges, the performance of the 2D harmonic retrieval is also improved by the proposed method. Moreover, the more accurate the prior knowledge, the better the performance of the proposed method.

\section{Conclusions}

In this paper, we study the problem of estimating MD frequency components of a spectrally sparse signal using the prior knowledge of the frequency intervals. We first formulate an FS atomic norm minimization problem for MD harmonic retrieval. Then, the MD-FS Vandermonde decomposition of block Toeplitz matrices on given intervals is proposed. It is shown that by using the MD-FS Vandermonde decomposition, we can convert the FS atomic norm minimization into semidefinite program. Numerical simulation results show that when prior knowledge is known, the proposed method can achieve significantly better performance than the traditional atomic norm approaches. And the more accurate the prior knowledge, the better the performance of the proposed method.

\bibliographystyle{IEEEtran}
\bibliography{database} 

\end{document}